\documentclass{aa}

\usepackage[varg]{txfonts}
\usepackage{graphicx}
\usepackage{natbib,twoopt}
\usepackage[breaklinks=true]{hyperref} 
\bibpunct{(}{)}{;}{a}{}{,}             

\newcommand{\bb}{^{\rm 3D}\textsc{Barolo}}

\newcommand{\htwo} {{\rm H}_{\rm 2}}

\newcommand{\cii} {{\rm [C}\,{\sc\rm II}{\rm ]}}
\newcommand{\ciifine} {[{\rm C}\,{\sc\rm II}]~158\!\operatorname{-}\!\!\mum}

\newcommand{\lya} {{\rm Ly}\alpha}

\newcommand{\kms} {\,{\rm km\,s}^{-1}}

\newcommand{\ergs} {\,{\rm erg\,s}^{-1}}
\newcommand{\erg} {\,{\rm erg}}

\newcommand{\mm} {\,{\rm mm}}
\newcommand{\mum} {\,\mu{\rm m}}

\newcommand{\de}{^{\circ}}

\newcommand{\pc} {\,{\rm pc}}

\newcommand{\kpct} {{\rm kpc}}
\newcommand{\kpc} {\,{\rm kpc}}

\newcommand{\mo}{\,{M}_\odot}

\newcommand{\Jy} {\,{\rm Jy}}
\newcommand{\mJy} {\,{\rm mJy}}

\newcommand{\mjybeam} {\,{\rm mJy}\,{\rm beam}^{-1}}

\newcommand{\yrinv}{\,{\rm yr}^{-1}}

\newcommand{\Myr}{\,{\rm Myr}}

\newcommand{\moyr}{\,{{M}_{\odot}\,\rm yr}^{-1}}

\newcommand{\gsim}{\lower.7ex\hbox{$\;\stackrel{\textstyle>}{\sim}\;$}}
\newcommand{\lsim}{\lower.7ex\hbox{$\;\stackrel{\textstyle<}{\sim}\;$}}

\newcommand{\vrot}{v_{\rm rot}}

\newcommand{\Ma}{{M}}

\newcommand{\Ms}{{M}_{\star}}
\newcommand{\Mg}{{M}_{\rm gas}}

\newcommand{\Mdyn}{{M}_{\rm dyn}}
\newcommand{\Mhtwo}{{M}_{\rm H_2}}

\newcommand{\Rd}{{R}_{\rm d}}
\newcommand{\Reff}{{R}_{\rm e}}

\newcommand{\sigmag}{\sigma_{\rm gas}}
\newcommand{\SFR}{\rm SFR}

\newcommand{\aCO}{\alpha_{\rm CO}}

\newcommand{\Ktab}{{\rm K}}

\newcommand{\cotwoone}{{\rm CO}\,({\rm J}=2\to1)}
\newcommand{\cofourthree}{{\rm CO}\,({\rm J}=4\to3)}
\newcommand{\cofivefour}{{\rm CO}\,({\rm J}=5\to4)}

\begin{document}

\title{Fast rotating and low-turbulence discs at $z\simeq 4.5$: Dynamical evidence of their evolution into local early-type galaxies}

\author{F.\ Fraternali
  \inst{1}
  \and
  A.\ Karim
  \inst{2}
  \and
  B.\ Magnelli
  \inst{2}
  \and
  C.\ G{\'o}mez-Guijarro
  \inst{3}
  \and
  E.\ F.\ Jim{\'e}nez-Andrade
  \inst{4}
  \and
  A.\ C.\ Posses
  \inst{5}
}
\institute{
  \inst{1} Kapteyn Astronomical Institute, University of Groningen, Postbus 800, 9700 AV Groningen, The Netherlands,
  \email{fraternali@astro.rug.nl} \\
  \inst{2} Argelander Institut f\"ur Astronomie, Universit\"at Bonn, Auf dem H\"ugel 71, Bonn, D-53121, Germany.\\ 
  \inst{3} AIM, CEA, CNRS, Universit{\'e} Paris-Saclay, Universit{\'e} Paris Diderot, Sorbonne Paris Cit{\'e}, F-91191 Gif-sur-Yvette, France.\\
  \inst{4} National Radio Astronomy Observatory, 520 Edgemont Road, Charlottesville, VA 22903, USA.\\
  \inst{5} N{\'u}cleo de Astronom{\'i}a, Facultad de Ingenier{\'i}a y Ciencias, Universidad Diego Portales, Av.\ Ej{\'e}rcito 441, Santiago, Chile.
  }
\date{Received ; accepted }


\abstract
    {Massive starburst galaxies in the early Universe are estimated to have depletion times of $\sim 100\Myr$ and thus be able to convert their gas very quickly into stars, possibly leading to a rapid quenching of their star formation.
      For these reasons, they are considered progenitors of massive early-type galaxies (ETGs).
      In this paper, we study two high-$z$ starbursts, AzTEC/C159 ($z\simeq 4.57$) and J1000+0234 ($z\simeq 4.54$), observed with ALMA in the $\ciifine$ emission line.
      These observations reveal two massive and regularly rotating gaseous discs.
      A 3D modelling of these discs returns rotation velocities of about $500 \kms$ and gas velocity dispersions as low as $\approx 20 \kms$, leading to very high ratios between regular and random motion ($V/\sigma \gsim 20$), at least in AzTEC/C159.
      The mass decompositions of the rotation curves show that both galaxies are highly baryon-dominated with gas masses of $\approx 10^{11}\mo$, which, for J1000+0234, is significantly higher than previous estimates.
      We show that these high-$z$ galaxies overlap with $z=0$ massive ETGs in the ETG analogue of the stellar-mass Tully-Fisher relation once their gas is converted into stars.
      This provides dynamical evidence of the connection between massive high-$z$ starbursts and ETGs, although the transformation mechanism from fast rotating to nearly pressure-supported systems remains unclear.
    }

\titlerunning{Fast rotating and low-turbulence discs at $z\simeq 4.5$}
\authorrunning{Fraternali et al.}

\maketitle

\section{Introduction}

Over the last decade, we have witnessed the accumulation of stunning new data that are changing our view of very distant galaxies down to the early epochs of their formation.
The dynamics of high-$z$ galaxies can be very efficiently studied using a variety of emission lines.
Optical recombination and forbidden lines are mostly used in the redshift range $0\lesssim z \lesssim 3.5$ \citep[e.g.][]{ForsterSchreiber+2009, Stott+2016, Turner+2017}, while at $z\gtrsim4$ the main contribution has come from CO and $\ciifine$ data \cite[e.g.][]{Carilli&Walter2013}.

One of the most surprising results to have emerged is the ubiquitous presence of rotating gaseous discs in high-$z$ galaxies \citep[e.g.][]{Genzel+2006}.
Initial estimates of the amount of rotation-dominated galaxies of around $30\%$ \citep{ForsterSchreiber+2009} have been significantly revised to the point that the dynamics of the vast majority of galaxies at $z\approx 1-5$ is now considered to be dominated by rotation \citep{Wisnioski+2019, Lelli+2018, Loiacono+2019, Neeleman+2020, Rizzo+2020, Lelli+2021}.
This revision has been the consequence of major improvements in the quality of the data and data analysis.
Kinematic data of high-$z$ galaxies suffer two main shortcomings: low angular resolutions and a low signal-to-noise ratio (S/N).
Both, however, have improved in recent years thanks, for instance, to the use of adaptive optics (AO) techniques in the optical \citep{ForsterSchreiber+2018} and the progressive addition of new antennas to the Atacama Large Millimeter/submillimeter Array (ALMA) interferometer.

The second advance of recent years has been the development of new software that allows the extraction of the entirety of the information contained in data cubes \citep[e.g.][]{DiTeodoro&Fraternali2015, Bouche+2015}.
The kinematics of disc (rotation-dominated) galaxies is studied by modelling the galaxy as a series of concentric rings characterised by different geometrical and kinematic parameters \citep[tilted-ring model;][]{Rogstad+1974}.
Emission-line velocity fields can be easily fitted with a tilted-ring model to extract the rotation curve when the galaxy is observed at high angular resolution, as what happens for nearby systems \citep[e.g.][]{Battaglia+2006, deBlok+2008}.
However, as the angular resolution decreases, line profiles from different parts of the disc, characterised by different line-of-sight velocities, blend together in the same resolution element, an effect known as `beam smearing' \citep{Swaters+2009}.
The best way to correct for this observational bias is to build artificial observations (3D tilted-ring models) and compared them directly with the emission in the data cube, without extracting 2D maps.
A public code that carries out this type of analysis is $\bb$ \citep{DiTeodoro&Fraternali2015}, which we use in this work.

This paper focuses on two starburst galaxies at $z\approx4.5$ observed with ALMA in the $\ciifine$ emission line.
Starburst high-$z$ galaxies are characterised by intense emission at wavelengths from far-infrared (FIR) to millimetre due to the re-processing of the ultraviolet (UV) light produced by young stars \citep[e.g.][]{Casey+2014}.
A sub-class of these sources, selected for their particularly strong sub-millimetre emission, are called sub-millimetre galaxies (SMGs) \citep{Weiss+2009, Karim+2013}.
The typical star formation rate (SFR) of these startbursts is of the order of hundreds of $\moyr$ \citep{daCunha+2015}, though exceeding $1000 \moyr$ in some cases \citep{Magnelli+2012, Swinbank+2014, Jimenez-Andrade+2020}.
These galaxies also have very high stellar masses, $\Ms\sim 10^{11}\mo$, leading to a specific star formation rate (sSFR) of$\,\sim 10^{-8}\yrinv$, which is characteristic of systems that are building their stellar component extremely rapidly \citep{Gomez+2018, Lang+2019, Dudzeviciute+2020}.
Estimates of the gas mass in massive high-$z$ starbursts return values comparable with their stellar mass; this gas is thought to be mostly in molecular form \citep{Ivison+2010, Harrington+2018}.
The depletion times ($M_{\rm H_2}/{\rm SFR}$) are very short ($\sim 100 \Myr$), again pointing at a rapid transformation of their gas reservoir into stars \citep{Jimenez-Andrade+2020, Birkin+2020}.

The environment of high-$z$ starbursts is thought to be violent and highly dynamic.
On the one hand, the extreme star formation activity in the galaxy is likely the consequence of a major merger or powerful gas accretion from the surroundings \citep{Riechers+2011, Aguirre+2013, Jimenez-Andrade+2020}.
On the other, star formation at these extraordinary rates is expected to cause intense supernova feedback that can eject a large portion of the gas \citep[e.g.][]{Hayward&Hopkins2017}.
Active galactic nuclei (AGNs) are also observed in high-$z$ starbursts, providing additional energy injection into the surrounding gas \citep[e.g.][]{Zeballos+2018}.

The kinematics of galaxies at $z\gsim 4$ has, until recent years, been challenging to study.
Rest-frame optical emission lines cannot be used at these high redshifts, while the $\lya$ line, due to resonant scattering, does not constitute a reliable kinematic tracer \citep[e.g.][]{Cantalupo+2019}.
Before the advent of ALMA, pioneering works provided indications of high rotation velocities using CO emission lines \citep{Greve+2005}. 
Observations with the Very Large Array (VLA) have given the first clear resolved observation of a regularly rotating disc in the SMG GN20 \citep{Hodge+2012}.
These authors reported a CO disc extending to a radius of about $7\kpc$ and with a rotation exceeding $500\kms$, assuming an inclination of the disc of $30\de$.
The dynamics of high-$z$ galaxies and AGNs has also been studied with the $\ciifine$ line, which is the brightest atomic fine-structure line in the interstellar medium (ISM) and is thought to be a good tracer of both molecular and atomic gas \citep{Carilli&Walter2013}.
This field has been revolutionised by ALMA, which led to the detection of several systems displaying regular gas kinematics \citep{Carniani+2013, DeBreuck+2014, Oteo+2016, Jones+2017}.
These galaxies tend to show very high rotation velocities, a signature of a population of massive objects, but less massive discs are also coming into reach \citep{Neeleman+2020, Rizzo+2020}.

Massive high-$z$ starbursts, and SMGs in particular, are thought to be progenitors of present-day (passive) early-type galaxies (ETGs) \citep{Cimatti+2008, Fu+2013, Toft+2014, Gomez+2018}.
Indeed, due to their extremely high SFRs, they form most of their stellar component at $z>2$ and are expected to have a sharp decline in their star formation history thereafter.
This hypothesis can also be tested dynamically, by investigating the shape of their gravitation potential.
Here we present a dynamical study of two starburst galaxies at $z\approx 4.5$, AzTEC/C159 and J1000+0234, observed with ALMA in the $\ciifine$ transition.
Our 3D approach reveals the existence of fast rotating and remarkably low-turbulence discs in these galaxies.
Section \ref{sec:data&modelling} presents the physical properties of our two galaxies.
Section \ref{sec:results} describes the 3D kinematic modelling of both galaxies and our main results, while Sect. \ref{sec:discussion} discusses the reliability of our measures and the possible future evolution of our starbursts.
Section \ref{sec:conclusions} sums up.
Throughout the paper we use Planck cosmological parameters \citep{Planck+2018}.

\section{Galaxy and data presentation}
\label{sec:data&modelling}

We present results for two galaxies at $z\approx 4.5$, AzTEC/C159 and J1000+0234, observed with ALMA in the $\ciifine$ fine-structure transition.
AzTEC/C159 was first detected through the ASTE/AzTEC-COSMOS 1.1-$\mm$ survey \citep{Aretxaga+2011}.
It is located at $z\simeq 4.57$ and has an estimated $\SFR\approx 700 \moyr$ and a stellar mass $\Ms\simeq 4.5\pm0.4 \times 10^{10}\mo$ \citep{Smolcic+2015,Gomez+2018}.
Low angular resolution observations in $\cotwoone$ and $\cofivefour$ revealed a tentative double-peaked profile consistent with a rotating disc \citep{Jimenez+2018}.
The total mass of molecular gas is $\Ma_{\htwo}\simeq 1.5\pm 0.3 \times 10^{11}\mo$, estimated using an $\aCO$ factor of $4.3 \mo (\Ktab \kms \pc^2)^{-1}$ \citep[see details in][]{Jimenez+2018}.
\citet{Gomez+2018} also estimated an effective radius of $\Reff\simeq 460^{+60}_{-240} \pc$ at rest-frame FIR wavelengths, which is indicative of a very compact object.
The galaxy is not detected by the Hubble Space Telescope ({\em HST}) in the F814W filter and was unfortunately not observed in the infrared (IR) filters \citep{Gomez+2018}.
We report the main physical properties of AzTEC/C159 in Table \ref{tab:aztec}.

\begin{table}[ht]
\caption{Physical properties of AzTEC/C159.}
\label{tab:aztec}      
\centering
\begin{tabular}{l c c}
\hline\hline
Property & Value & Ref \\
\hline                        
Kinematic centre (R.A.) & $9^{\rm h}$\,$59^{\rm m}$\,$30.401^{\rm s}\pm 0.003^{\rm s}$ & 1\\
Kinematic centre (Dec.) & $+1{\de}$\,$55'$\,$27.57''\pm 0.05''$ & 1\\
$z$   & $4.5668\pm 0.0002$ & 1\\
Stellar mass ($\Ms$) &  $4.5 \pm 0.4 \times 10^{10} \mo$ & 2\\
FIR effective radius ($\Reff$) & $460^{+60}_{-240} \pc$ & 2\\
Star form.\ rate (SFR) & $740^{+210}_{-170} \moyr$& 3\\
Mol.\ gas mass ($\Mhtwo$) & $1.5 \pm 0.3 \times 10^{11}\mo$ & 3\\
Disc inclination ($i$) & $40 \pm 10\de $ & 1\\
Disc position angle (PA) & $170 \pm 10\de$ & 1\\
$\ciifine$ flux  & $4.6\pm 0.5 \Jy\kms$ & 1\\
Rotation velocity ($\vrot$) &  $506^{+151}_{-76} \kms$ & 1\\
Velocity dispersion ($\sigmag$) & $16\pm 13\kms$ & 1\\
Conversion $''$/$\kpc$ & 6.70 & \\
\hline
\hline
\end{tabular}
\tablebib{
   (1) This work; (2)~\citet{Gomez+2018}; (3) \citet{Jimenez+2018}.
}
\end{table}

J1000+0234 (also known as AzTEC/C17) was selected as a $V$-band dropout in the COSMOS field and appears as a 3-$\sigma$ detection in JCMT/AzTEC-COSMOS.
It is located at $z\simeq4.54$ and has an IR-estimated SFR of $440\moyr$ with large uncertainties (Table \ref{tab:j1000}).
Its stellar mass is also rather uncertain: \citet{Smolcic+2015} reported a value of $8.7\times 10^{10}\mo$, while \citet{Gomez+2018} obtained $1.4\times 10^{10} \mo$.
Both authors fit the spectral energy distribution (SED) of J1000+0234 but used different techniques and considered different sizes for the actual source \citep[see][for details]{Gomez+2018}.
J1000+0234 appears in {\em HST} images as being composed of two sources with confirmed similar redshifts \citep{Capak+2008}.
These are visible in the top left panel of Fig.\ \ref{fig:j1000}: The first is clearly associated with the $\cii$ emission, while the second is located in the south and is significantly outside the region of the line emission.
The source on the west side of the {\em HST} image is a foreground galaxy.
Because of this peculiar configuration and the large velocity offset revealed by the $\lya$ emission \citep{Capak+2008}, J1000+0234 has been classified as a merger.
Table \ref{tab:j1000} summarises the physical properties of this galaxy.

With 870-$\mum$ flux densities of $\approx 7 \mJy$, as measured from the continuum emission observed in the line-free band-7 channels, both AzTEC/C159 and J1000+0234 form part of the unblended bright field population of sub-millimetre sources \citep[e.g.][]{Karim+2013, Stach+2018, Simpson+2020}.
Given that the bulk of these SMGs that have been interferometrically detected at 870-$\mum$ flux densities above 6 mJy reside at much lower redshifts \citep[$z_{\rm med} = 2.9$,][]{Simpson+2020} compared to our two galaxies, they belong to a rather rare sub-class of the broader SMG population.
In the following, we refer to these two galaxies as starbursts \citep[see also][]{Gomez+2018} given that the available constraints for their stellar mass content and ongoing star formation activity places them above the $z\gsim 4$ main sequence of star forming galaxies \citep{Steinhardt+2014, Caputi+2017, Iyer+2018}.
However, it should be noted that more robust constraints, in particular of their stellar mass budget and spatial distribution (e.g.\ through observations with the James Webb Space Telescope), are needed to clarify their starburst nature.

\begin{table}[ht]
\caption{Physical properties of J1000+0234.}
\label{tab:j1000}      
\centering
\begin{tabular}{l c c}
\hline\hline
Property & Value & Ref \\
\hline                        
Kinematic centre (R.A.) & $10^{\rm h}$\,$0^{\rm m}$\,$54.493^{\rm s}\pm 0.003^{\rm s}$ & 1\\
kinematic centre (Dec.) & $+2{\de}$\,$34'$\,$36.12''\pm 0.05''$  & 1\\
$z$   & $4.5391\pm 0.0004$ & 1\\ 
Stellar mass ($\Ms$) &  $(1.4-8.6) \times 10^{10} \mo$ & 2, 3\\
FIR effective radius ($\Reff$) & $700^{+120}_{-100} \pc$ & 2\\
Star form.\ rate (SFR) & $440^{+1200}_{-320} \moyr$& 2\\
Mol.\ gas mass ($\Mhtwo$) & $(2.6-15) \times 10^{10}\mo$ & 4, 1 \\
Disc inclination ($i$) & $75 \pm 10\de $ & 1\\
Disc position angle (PA) & $145 \pm 5\de$ & 1\\
$\ciifine$ flux  & $5.9\pm 0.5 \Jy\kms$ & 1\\
Rotation velocity ($\vrot$) &  $538\pm40 \kms$ & 1\\
Velocity dispersion ($\sigmag$) & $\lsim 60 \kms$ & 1\\
Conversion $''$/$\kpct$ & 6.72 & \\
\hline
\hline
\end{tabular}
\tablebib{
   (1) This work; (2)~\citet{Gomez+2018}; (3) \citet{Smolcic+2015}; (4) \citet{Schinnerer+2008}.
}
\end{table}

\begin{figure*}[ht]
\centering
  \includegraphics[width=\textwidth]{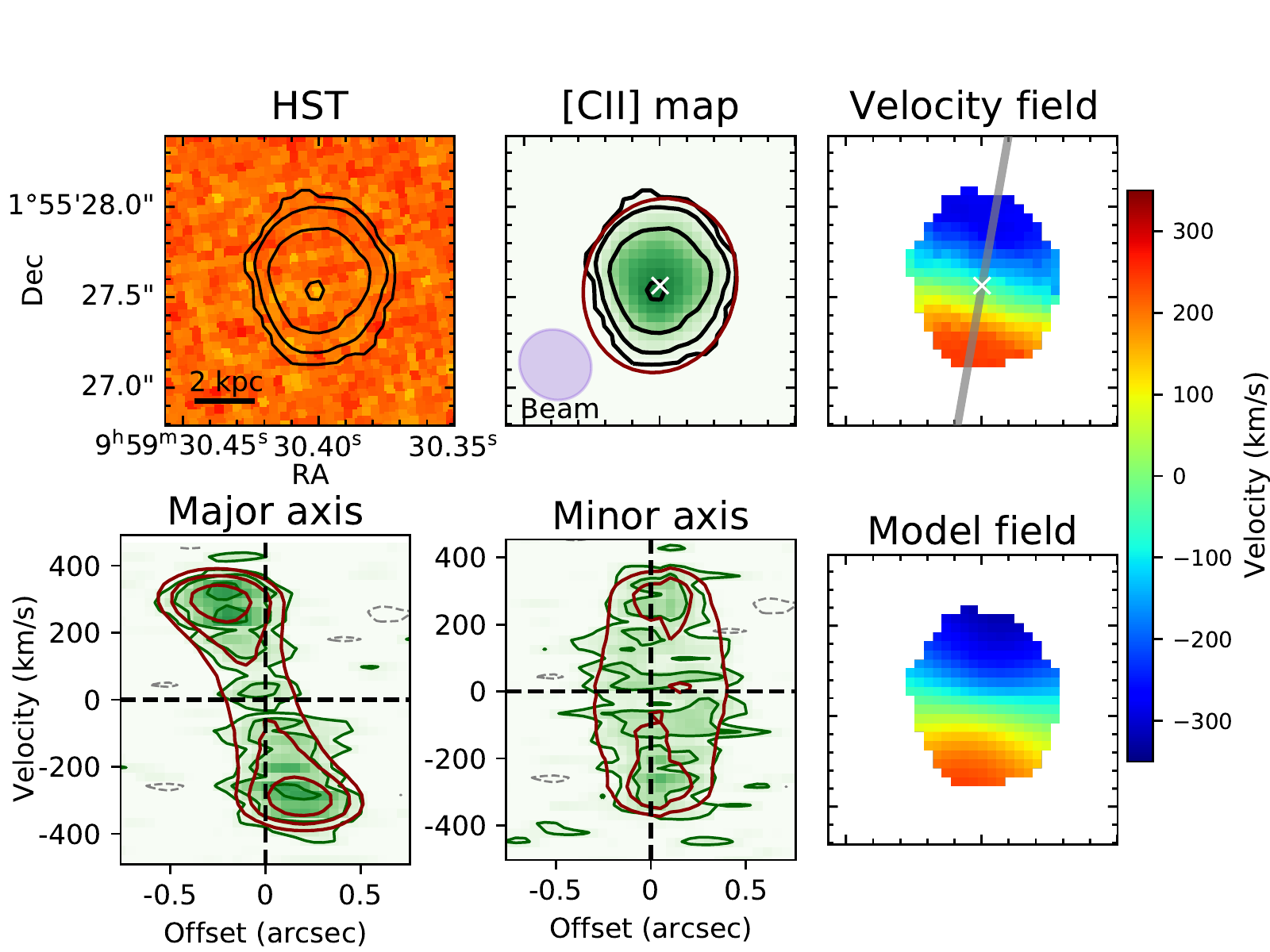}
  \caption{$\ciifine$ observations and modelling of AzTEC/C159. {\em Top left panel}: {\em HST} F814W image with the $\ciifine$ total-flux map contours overlaid; the source is not detected by {\em HST}.
    {\em Top middle panel}: Total-flux map (green shape and black contours) of the $\ciifine$ emission; the lowest contour is at about 2.5 times the r.m.s.\ noise in this map, and the other contours are at 2, 4, and 8 times this level. In red we show the outer contour of the total-flux map of the model obtained with $\bb$ (convoluted with the observational beam, shown in the bottom left corner).
    {\em Top right panel}: Velocity (weighted-mean) field of the $\ciifine$ line.
    {\em Bottom left panel}: Position-velocity diagram along the kinematic major axis of the galaxy (grey line across the velocity field); the data are in green shades and contours, while the best-fit $\bb$ model is in red contours. Contour levels are at $-2$, 2, 4, and 8 times the r.m.s.\ noise per channel.
    {\em Bottom centre panel}: Position-velocity diagram along the kinematic minor axis of the galaxy displayed analogously to the major-axis plot.
    {\em Bottom right panel}: Velocity (weighted-mean) field of the best-fitting $\bb$ model.
  }
  \label{fig:aztec}
\end{figure*}

Our two starburst galaxies were observed during ALMA Cycle 2 (program 2012.1.00978.S; PI: A. Karim) in June 2014.
They were part of an original sample of the only seven spectroscopically confirmed (Ly-$\alpha$ detection using Keck/DEIMOS) $z\gsim 4$ galaxies in the COSMOS field \citep[e.g.][]{Schinnerer+2008}.
The ALMA band 7 was used to cover the emission of $\ciifine$ at $z\approx 4.5$.
Only three galaxies of the original sample were eventually observed, the third being Vd-17871, which we do not present here as the data are not of sufficient quality for a reliable kinematic fit.
Standard data reduction was carried out using CASA; details can be found in \citet{Gomez+2018}.
Data cubes were produced using a standard Briggs robust parameter of 0.5 and 2, leading to slightly different angular resolutions (less than a $10\%$ difference).
For the present analysis, we used data cubes with robust=2 (which maximises the S/N) for AzTEC/C159 and robust=0.5 (a compromise between resolution and S/N) for J1000+0234.
For both sources, we produced data cubes with different channel separations.
The low spectral resolution data cube, with a channel separation of $\approx 25 \kms$, has a higher S/N per channel and gives the most reliable kinematic fit and a better comparison between data and models.
The high spectral resolution data, with a channel separation of $\approx 10 \kms$, were used to better constrain the (intrinsic) gas velocity dispersion in AzTEC/C159 (Sect. \ref{sec:fitAztecHR}).
We note that in this type of data, the channel separation can be considered equal to the velocity resolution (full width at half maximum) of the data.
The observational properties of the data cubes are given in Tables \ref{tab:aztecObs} and \ref{tab:j1000Obs}.

\section{Data analysis and results}
\label{sec:results}

We analysed our data with the software $\bb$ \citep{DiTeodoro&Fraternali2015} in order to characterise the source geometry and the kinematics of the gas.
In this section, we present the rotation curves and velocity dispersions obtained for both galaxies, first using the low spectral resolution (but high S/N) data cubes and then focusing on the velocity dispersion of AzTEC/C159 at high spectral resolution.
Throughout this paper, we use the term `velocity dispersion' to refer to the intrinsic value, corrected for instrumental effects (spatial and spectral resolution).
The non-corrected quantity is referred to as the `broadening' of the emission line.

\subsection{3D kinematic fit of AzTEC/C159}
\label{sec:fitAztec}

The total-flux map and velocity field of the $\ciifine$ emission of AzTEC/C159 obtained with the low spectral resolution data cube (Table \ref{tab:aztecObs}) are shown in the top panels of Fig.\ \ref{fig:aztec}.
The total-flux map (analogous to a zeroth moment) has been constructed by summing the emission across all channel maps after the application of a mask.
The mask is calculated by $\bb$ through smoothing and clipping the data cube, followed by a source detection in the 3D space \citep{DiTeodoro&Fraternali2015}.
After several experiments with either only smoothing and clipping or only source detection, we found that a combination of the two produces the best result of a mask that is not too tight around the emission but also not so large as to include too much noise.
The top left panel of Fig.\ \ref{fig:aztec} shows the total-flux map in contours\footnote{We note that, given that we are using a mask, the noise is not constant across the total-flux map, although we have verified that the variations along the outer contours shown in Figs.\ \ref{fig:aztec} and \ref{fig:j1000} are small (less than 20\%). For more details about the S/N levels of outer contours in these types of maps, see \citet{Iorio+2017}.} overlaid on an {\em HST} image at the location of AzTEC/C159; it shows no detection in the F814W filter, which roughly corresponds to rest-frame far-UV.
This lack of detection is likely due to the intense dust absorption \citep{Gomez+2018}.
The detection in IR with {\em Spitzer} Infrared Array Camera (IRAC) is compact and does not give us information about the morphology of this source \citep[see Fig.\ 1 in][]{Smolcic+2015}.

\begin{table}[ht]
\caption{$\cii$ observational properties of AzTEC/C159.}
\label{tab:aztecObs}      
\centering
\begin{tabular}{l l c c }
\hline\hline
Data cube & Property & Units & Value \\
\hline
Low-res \\
& Robust &  & 2 \\
& Beam size & $''$ & $0.38\times 0.35$ \\
& r.m.s.\ noise  & $\mjybeam$ & 0.8 \\
& Channel separation & $\kms$ & 27.4 \\
High-res & & &\\
& Robust &  & 2 \\
& Beam size & $''$ & $0.38\times 0.36$ \\
& r.m.s.\ noise  & $\mjybeam$ & 1.2 \\
& Channel separation & $\kms$ & 10.3 \\
\hline
\hline
\end{tabular}
\end{table}

\begin{figure*}[ht]
\centering
  \includegraphics[width=0.9\textwidth]{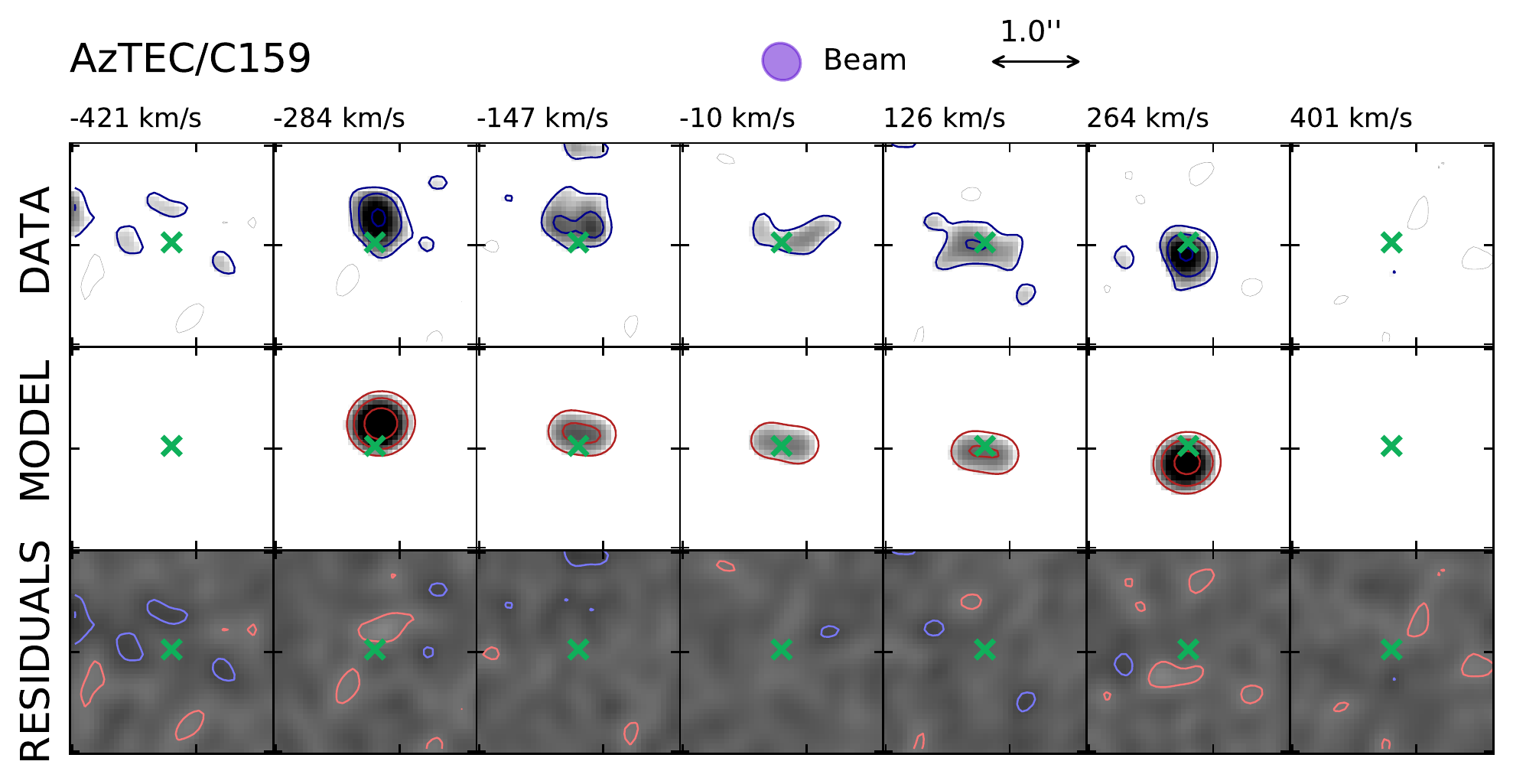}
  \caption{Comparison between data and $\bb$ best-fit model for AzTEC/C159. {\em Top row}: Seven representative channel maps showing the $\ciifine$ emission of AzTEC/C159 (low spectral resolution data cube; see Table \ref{tab:aztecObs}).
    The contours are at $-$2, 2, 4, and 8 times the r.m.s.\ noise per channel ($\sigma$).
    Line-of-sight velocities with respect to systemic velocity (set at $v=0$) are shown on top of each channel map.
    {\em Middle row}: Channel maps of the axi-symmetric best-fit model obtained with $\bb$; the contours are at the same levels as in the data.
    {\em Bottom row}: Residuals (data minus model); pink contours are negative ($-$2 $\sigma$), and light blue contours are positive ($+$2 $\sigma$).
    In all panels, the green cross shows the dynamical centre, whose coordinates are given in Table \ref{tab:aztec}.
  }
  \label{fig:aztecChannels}
\end{figure*}

The total-flux map of the $\cii$ line of AzTEC/C159 is clearly resolved, and it appears slightly elongated in, roughly, the north-south direction (we note that this is nearly orthogonal to the beam elongation).
The velocity field of the galaxy (top right panel of Fig.\ \ref{fig:aztec}) shows a clear gradient aligned with the geometrical elongation of the source, an indication that we are in the presence of a rotating disc.
The morphology of the velocity field is largely dominated by the beam smearing.
Therefore, the velocity field has not been used in the fit of the kinematics in any way, and it is only shown here as an illustration.
Figure \ref{fig:aztec} also presents two position-velocity (p-v) diagrams (green shade and contours) along the major and the minor kinematic axes.
The p-v diagram along the major axis shows an extremely clear pattern of rotation, with a considerable symmetry between approaching and receding sides.
The p-v diagram along the minor axis is instead typical of galaxies with a fast rising rotation curve seen at low angular resolution \citep[see Fig.\ 7 in][]{DiTeodoro&Fraternali2015}.

In order to fit the line emission from a galaxy, $\bb$ creates 3D realisations of tilted-ring models \citep{Rogstad+1974}.
In practice, this is a mock observation built as a kinematic model of concentric rings where the gas motion is characterised by 1) rotation and 2) random motions (that we ascribe to turbulence).
After producing the mock observation, the algorithm convolves it with the beam before calculating and eventually minimising the residuals between data and model.
This approach allows us to account for the beam smearing effect in the best possible way, that is,\ during the fitting process \citep{DiTeodoro+2016,Iorio+2017,ManceraPina+2020}. 
Our AzTEC/C159 data have very few resolution elements across the galaxy, and thus we chose to use only two rings.
These rings define two adjacent annuli, each with a radial width of $0.23''$.
Given the size of the beam (Table \ref{tab:aztecObs}), this choice makes the properties extracted from the two rings not fully independent.

The geometrical parameters of the rings were determined automatically by $\bb$. 
The centre of the rings was found by the source detection algorithm and confirmed by the kinematic fit (Table \ref{tab:aztec}).
The kinematic centre of AzTEC/C159 is about $0.15''$ away from the VLA radio continuum peak in \citet{Smolcic+2015}.
The systemic (central) frequency of the source, which corresponds to a systemic redshift, was determined by $\bb$ using the global $\cii$ profile \citep[see Fig.\ 2 in][]{Jimenez+2018} and was refined by about $10\kms$ following an inspection of the p-v diagrams.
The position angle of the disc is nicely constrained by the gradient in the velocity field (the top right panel of Fig.\ \ref{fig:aztec}).

The most problematic parameter is the inclination of the disc along the line of sight.
This was estimated by trial and error and by comparing the outer contours of the model total-flux map with the total $\ciifine$ map.
Both the total-flux map and the channel maps (Fig.\ \ref{fig:aztecChannels}) clearly exclude a highly inclined disc, pointing to values between 30 and 50 degrees with the most likely value around $40\de$, which we take as our fiducial.
To test this trial-and-error result, we also computed the residuals between the total-flux maps of the $\bb$ models with varying inclinations and the observed map.
This technique was used in \citet{ManceraPina+2020}, who showed its good performance through intensive experiments, including with mock data obtained from hydrodynamic simulations.
The residuals that we found display a clear minimum at $42\de$, confirming the trial-and-error value.
The red ellipse overlaid in the top middle panel of Fig.\ \ref{fig:aztec} shows the outer contour of a model produced with an inclination of 40 degrees; we note how it follows the outer contour of the data almost in its entirety.
Given the uncertainty of dealing with a disc at a relatively low inclination, we chose to adopt a conservative error of $10 \de$ in the final value (Table \ref{tab:aztec}).
This error in inclination is the main contributor to the uncertainty in the rotation velocity for AzTEC/C159.

We finally ran $\bb$ keeping all the geometrical parameters and the systemic frequency fixed and simultaneously fitting the rotation velocity and the velocity dispersion of the gas.
The surface density of the model is derived directly from the data as an azimuthal average within the annuli.
In Fig.\ \ref{fig:aztec}, we overlay the contours from our best-fit model cube on the p-v diagrams along the major and minor axes, while in Fig.\ \ref{fig:aztecChannels} we show some representative channel maps of the data and model as well as the resulting residuals.
It is clear from these comparisons that the fit is of very high quality and that the $\cii$ emission of AzTEC/C159 is reproduced very well by a disc in differential rotation.

\begin{figure}[ht]
\centering
  \includegraphics[width=0.5\textwidth]{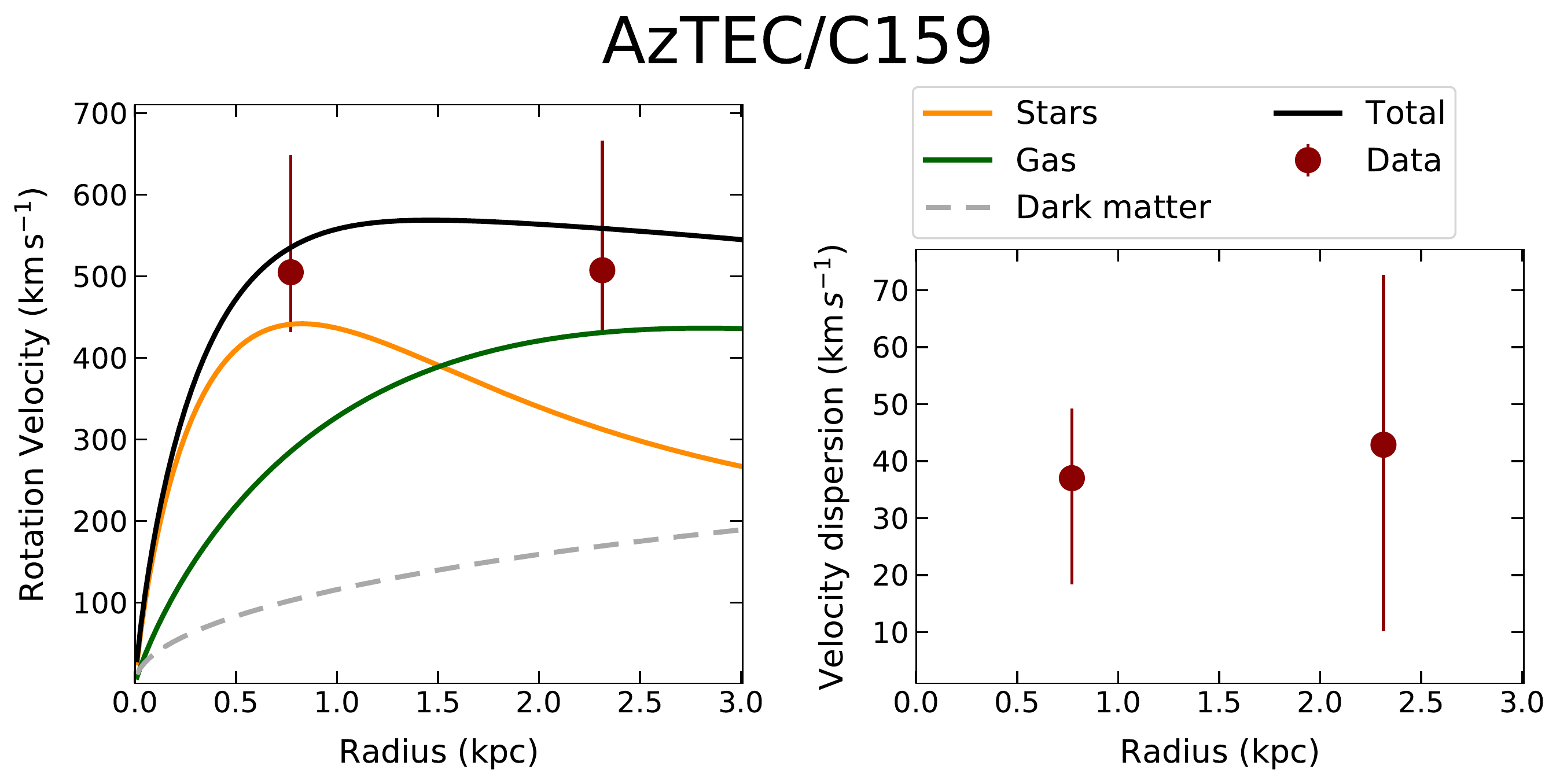}
  \caption{Best-fit rotation velocities and velocity dispersions of AzTEC/C159. {\em Left panel}: Rotation velocities of the two annuli in which we decomposed the disc of AzTEC/C159.
    The orange and green curves show the contribution of stars and gas, respectively, modelled as exponential discs, the dashed grey curve shows that of a dark matter halo modelled as an NFW profile, and the black curve is the sum of all components.
    This comparison is illustrative and not fitted to the data points.
    {\em Right panel}: Gas velocity dispersion in the two annuli obtained using the low spectral resolution data cube (see Sect. \ref{sec:fitAztecHR} for the high-resolution values).
  }
  \label{fig:aztecVrotVdisp}
\end{figure}

\subsection{Rotation curve analysis of AzTEC/C159}
\label{sec:rcAztec}

The values of rotation velocity and velocity dispersions in the two rings are shown in Fig.\ \ref{fig:aztecVrotVdisp}.
In order to calculate the errors, $\bb$ performs a Monte Carlo sampling of the parameter space around the best-fit value.
However, this does not take into account the uncertainties on parameters that have been fixed in the fit.
As mentioned, the main uncertainty for AzTEC/C159 is given by the relatively low inclination angle of the disc.
Therefore, we calculated the errors by re-running $\bb$ and fixing the inclination first to $30\de$ (the upper bound in rotation velocity) and then to $50\de$ (the lower bound).
The resulting errors were summed in quadrature with the nominal (Monte Carlo) errors of the fit, which in any case are subdominant.
For consistency, we used the same method to estimate the errors in the velocity dispersion.

The two annuli return rotation velocities and velocity dispersions consistent with each other.
Regarding the rotation velocity, this result hints at a flat rotation curve, although we caution that our two rings are not fully independent and that it is difficult to draw conclusions about the precise shape of the curve.
The resulting rotation velocity, averaged between that of the two rings, is very high, $\vrot = 506^{+151}_{-76}\kms$, consistent with the value found by \citet{Jones+2017} if we use their inclination of $30\de$.
From this rotation velocity (assuming it is equal to the circular speed $v_{\rm c}$), we can derive a dynamical mass\footnote{This is calculated with the standard formula $M(<R_{\rm out})=R_{\rm out}v_{\rm c}^2(R_{\rm out})/G$ (with $R_{\rm out}$ the radius of the outer ring), which gives an estimate that is correct within 10\% for exponential discs at radii $R\gsim 2.2 \Rd$, i.e.\ around and beyond the peak of their contribution \citep{Binney&Tremaine2008}. The errors in the dynamical mass are calculated considering the 1-$\sigma$ uncertainty in $v_{\rm c}$.} out to the outer measured point of $\Mdyn=1.4^{+0.9}_{-0.4}\times 10^{11}\mo$.
As an illustration, in Fig.\ \ref{fig:aztecVrotVdisp} (left panel) we show the contribution of two razor-thin exponential discs \citep{Freeman1970} to represent the contribution of stars and gas.
For the stars, we used the FIR effective radius of $\Reff\simeq 460^{+60}_{-240} \pc$ determined at $\approx 870\mum$ (rest-frame $\approx 160\mum$) by \citet{Gomez+2018}.
We converted this FIR radius to an approximate rest-frame optical effective radius using a multiplicative factor of 1.4 \citep{Fujimoto+2017}.
From this we derived the exponential scale-length $R_{\rm d}$ using the standard conversion $R_{\rm d}=R_{\rm e}/1.678$.
Such a stellar component, with the measured stellar mass\footnote{This stellar mass, as well as the gas mass below, are used to derive the central surface brightness of the exponential disc as $\Ms/(2\pi \Rd^2)$.} of $4.5\times10^{10}\mo$ (Table \ref{tab:aztec}), gives a contribution that appears to reproduce the high rotation velocity close to the centre of the galaxy well (first point of the rotation curve).

We derived the exponential scale-length of the gas disc from the $\cii$ zeroth moment map using Galfit \citep{Peng+2010a} and obtained $R_{\rm gas}=1.3\pm 0.4 \kpc$.
This, combined with a gas mass of $1.5\times 10^{11} \mo$ \citep{Jimenez+2018}, gives the contribution shown by the green curve in Fig.\ \ref{fig:aztecVrotVdisp}.
For the dark matter halo, we assumed a Navarro-Frenk-White (NFW) profile \citep{Navarro+1996} with a concentration $c_{\rm 200}=3.4$, which is appropriate for halos of very different masses at these redshifts \citep{Dutton&Maccio2014}, and a mass $M_{200}=2\times 10^{12}\mo$ \citep[e.g.][]{Rizzo+2020}.
Despite the large mass, it is clear that the dark matter remains subdominant (see dashed grey curve) and only completely unrealistic values of $M_{200}>10^{14}\mo$ would give a contribution in these inner regions of the galaxy comparable to the baryons.
Using alternative profiles for the dark matter halo (such as the pseudo-isothermal profile) also makes only a negligible difference.
Unfortunately, the low angular resolution of our data does not allow for a full decomposition of the rotation curve \citep{Cimatti+2019}; however, we can derive two important indications from this analysis.
First, it is clear that the dynamical properties of AzTEC/C159 show that we are dealing with an extremely baryon-dominated object (the baryon-to-dark mass ratio at the outer measured point is $\approx 10$).
Second, the available estimates of masses and sizes of both the stellar and gaseous discs are in good agreement with our inferred dynamical properties.

The velocity dispersion in Fig.\ \ref{fig:aztecVrotVdisp} (right panel) shows large error bars but favours values of the order of a few tens of $\kms$.
We note that this value is close to the velocity resolution of our data, meaning that the velocity dispersion cannot be reliably measured by $\bb$. 
Clearly, this points to a strongly rotation-dominated system, with values of $V/\sigma \gsim 10$.
We return to this point in Sect. \ref{sec:fitAztecHR}.

\subsection{3D kinematic fit of J1000+0234}
\label{sec:fitJ1000}

\begin{figure*}
\centering
  \includegraphics[width=\textwidth]{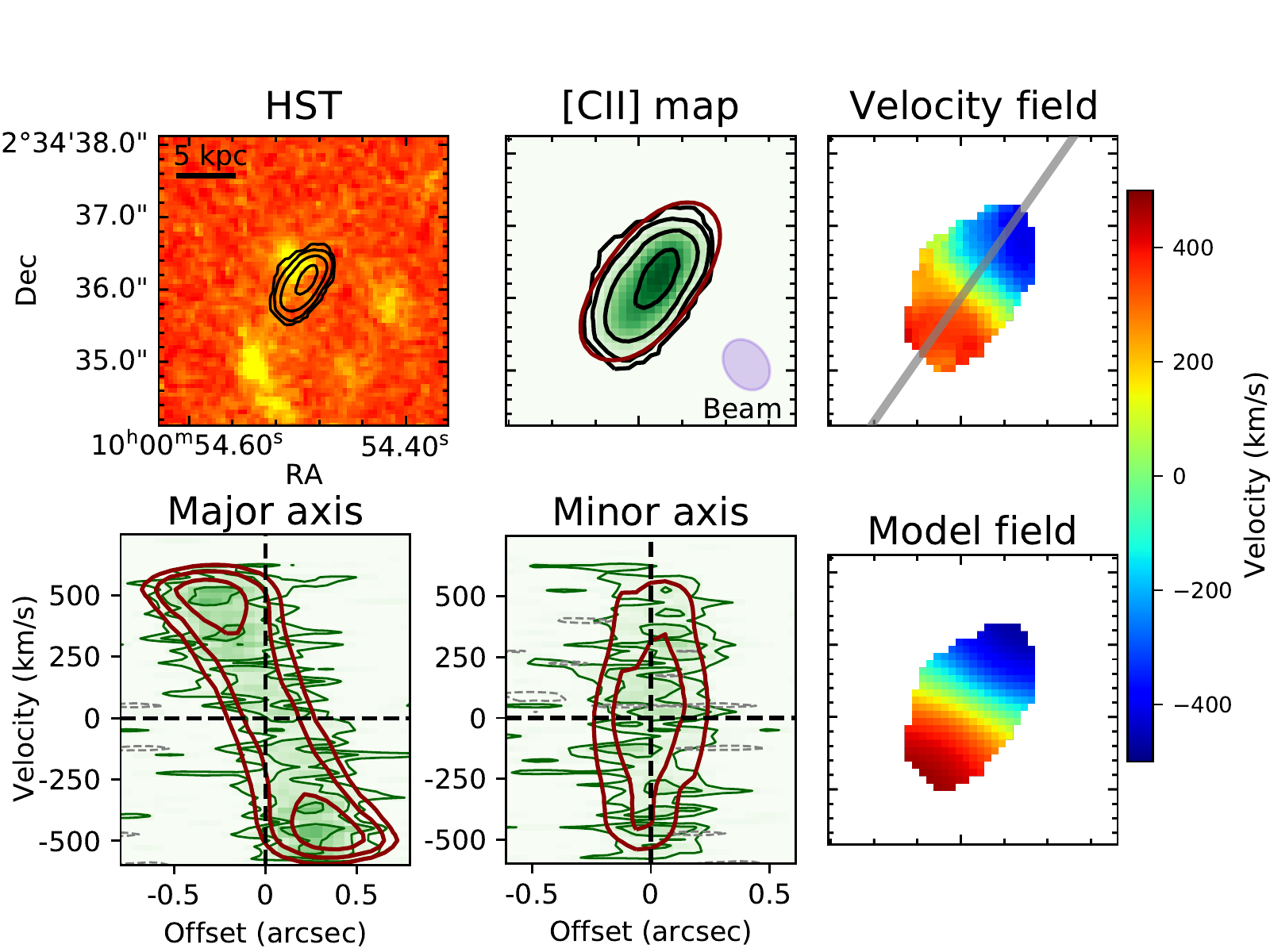}
  \caption{$\ciifine$ observations and modelling of J1000+0234. {\em Top left panel}: {\em HST} WFC3 F160W image with the $\ciifine$ total-flux map contours overlaid.
    {\em Top middle panel}: Total-flux map (green shapes and black contours) of the $\cii$ emission; the lowest contour is at about 3 times the r.m.s.\ noise in this map, and the other contours are at 2, 4, and 8 times this level. In red we show the outer contour of the total-flux map of the best-fit model obtained with $\bb$ (convoluted with the observational beam, shown in the bottom right corner).
    {\em Top right panel}: Velocity (weighted-mean) field of the $\ciifine$ line.
    {\em Bottom left panel}: Position-velocity diagram along the kinematic major axis of the galaxy (grey line overlaid on the velocity field); the data are in green shades and contours, while the $\bb$ model is in red contours. Contour levels are at $-$2, 2, 4, and 8 times the r.m.s.\ noise per channel.
    {\em Bottom centre panel}: Position-velocity diagram along the kinematic minor axis of the galaxy displayed analogously to the major-axis plot.
    {\em Bottom right panel}: Velocity (weighted-mean) field of the best-fit $\bb$ model.
  }
  \label{fig:j1000}
\end{figure*}

In this section, we describe the data of our second galaxy, J1000+0234, and the kinematic fit that we performed with $\bb$ to obtain the rotation velocity and the velocity dispersion.
A number of descriptions are similar to those in Sect. \ref{sec:fitAztec} and are thus only briefly repeated.

The total-flux map of the $\ciifine$ emission of J1000+0234 is shown in the top (left and middle) panels of Fig.\ \ref{fig:j1000}.
Contrary to AzTEC/C159, this galaxy appears very elongated in $\cii$ emission, with an axis ratio of roughly two despite the low angular resolution.
It should be noted that the ALMA beam is elongated in the opposite direction. 
Thus, the $\cii$ morphology points to a highly inclined (nearly edge-on) galaxy.
As mentioned, {\em HST} shows a complex structure characterised by two emission regions, one of which overlaps with the $\cii$ emission although not with the peak of this emission.
This again can be due to dust absorption, and, as a consequence, the morphology of the stellar component in J1000+0234 is largely unconstrained.
\begin{table}[ht]
\caption{$\cii$ observational properties of J1000+0234.}
\label{tab:j1000Obs}      
\centering
\begin{tabular}{l  c c }
\hline\hline
Property & Units & Value \\
\hline
 Robust &  & 0.5 \\
 Beam size & $''$ & $0.38\times 0.29$ \\
 r.m.s.\ noise & $\mjybeam$ & 0.53 \\
 Channel separation & $\kms$ & 25.0 \\
\hline
\hline
\end{tabular}
\end{table}

The velocity field of J1000+0234 (top right panel in Fig.\ \ref{fig:j1000}) shows a regular gradient along the elongation of the emission, suggesting rotation.
This is corroborated by the p-v diagram along the major axis of the emission shown in the bottom left panel.
As for AzTEC/C159, we see here the typical pattern of a rotating disc observed at low angular resolution \citep{DiTeodoro&Fraternali2015}.
The p-v diagram along the minor axis again shows a very broad emission indicative of a very fast rising rotation curve combined with a strong beam smearing effect.
As already noted by \citet{Schinnerer+2008} for CO, the global extent of the emission line exceeds $1000\kms$.
In the following, we show that this is the result of the very fast (differential) rotation of the gaseous disc.

The fitting of J1000+0234 with $\bb$ is somewhat easier than that of AzTEC/C159 given the high inclination of the disc, which makes the final value of the rotation velocity very robust.
Here, we have assumed an inclination of the disc of $75\de$.
This value was obtained using trial and error and by comparing the model total-flux map with that of the data; the red contour in the top middle panel of Fig.\ \ref{fig:j1000} shows that a model with this inclination reproduces the $\cii$ outer contour very well.
The calculation of the residuals between total-flux maps of $\bb$ models with varying inclinations and the observed total-flux map returns a minimum at $73\de$, confirming the trial-and-error value.

\begin{figure*}[ht]
\centering
  \includegraphics[width=0.9\textwidth]{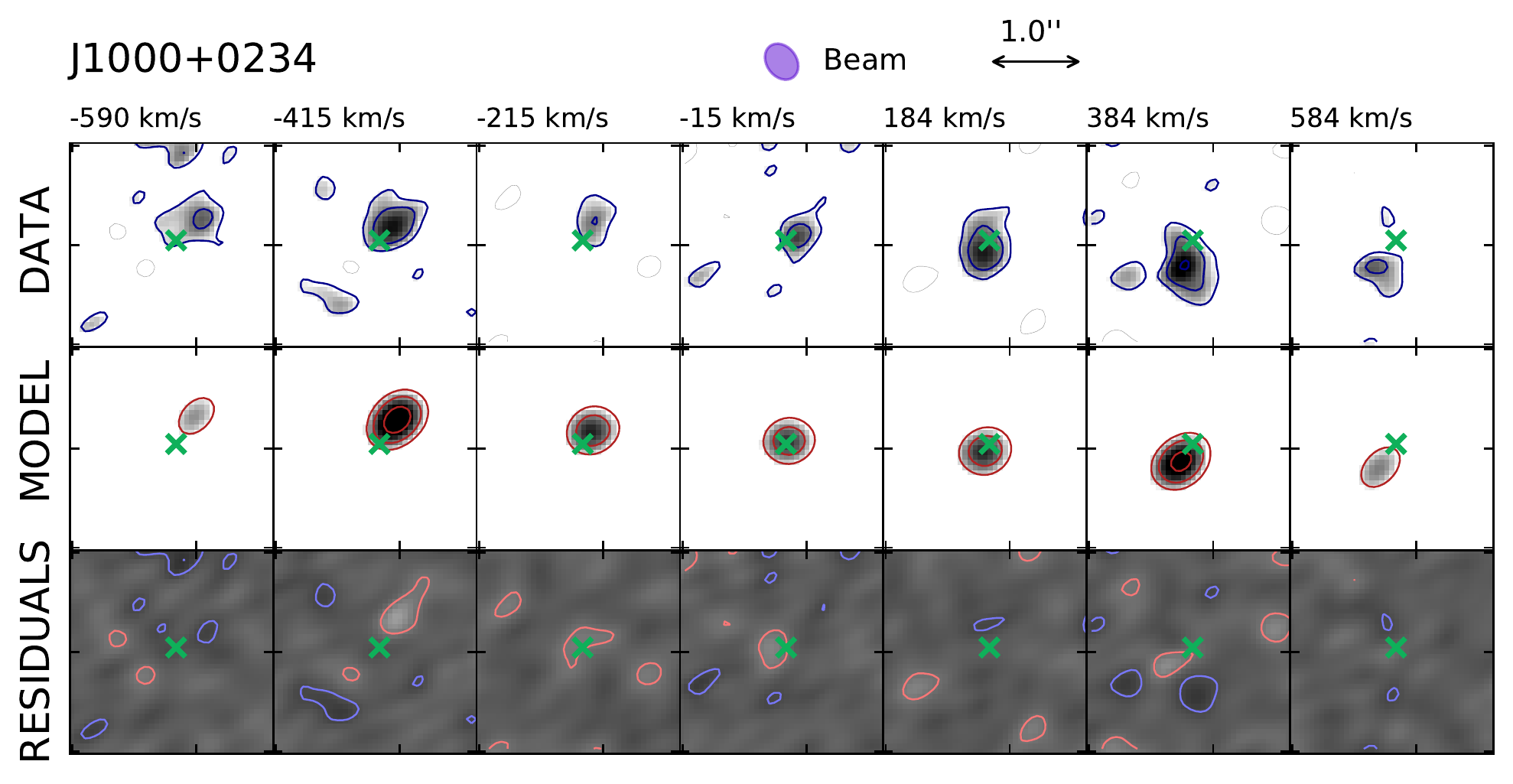}
  \caption{Comparison between data and $\bb$ best-fit model for J1000+0234. {\em Top row}: Seven representative channel maps showing the $\ciifine$ emission of J1000+0234.
    The contours are at $-$2, 2, 4, and 8 times the r.m.s.\ noise per channel ($\sigma$).
    Line-of-sight velocities with respect to the systemic velocity (set at $v=0$) are shown on top of each channel map.
    {\em Middle row}: Channel maps of the axi-symmetric best-fit model obtained with $\bb$; the contours are at the same levels as in the data.
    {\em Bottom row}: Residuals (data$-$model); pink contours are negative ($-$2 $\sigma$), and light blue contours are positive ($+$2 $\sigma$).
    In all panels, the green cross shows the dynamical centre, whose coordinates are given in Table \ref{tab:j1000}.
  }
  \label{fig:j1000Channels}
\end{figure*}

The other parameters of the disc were determined in a way analogous to AzTEC/C159. 
Small adjustments with respect to $\bb$ estimates were performed after visual inspection, but they were of the order of, or smaller than, the quoted errors (Table \ref{tab:j1000}).
The systemic frequency is somewhat uncertain because the emission is so broad that it reaches the edge of the ALMA band.
However, the shape of the channel maps (Fig.\ \ref{fig:j1000Channels}) and of the p-v diagrams in Fig.\ \ref{fig:j1000} indicate that the disc emission beyond the edge of the band must be very limited (at most a channel or two, assuming symmetry between the approaching and receding sides).
The kinematic centre that we found is $0.4''$ away from the peak of the $\cofourthree$ emission in \citet{Schinnerer+2008}, which was obtained from data with a resolution of roughly $2''$.

Once all the geometrical parameters were fixed (Table \ref{tab:j1000}), we ran $\bb$ to simultaneously fit rotation velocity and velocity dispersion.
Given the size of the galaxy along the major axis and the orientation of the beam, we used three adjacent annuli that were each $0.2''$ wide.
Regarding AzTEC/C159, this is a small oversampling of our data (see Table \ref{tab:j1000Obs}) that makes the three values that we obtained not fully independent.
However, it allows us to have an indication of the shape of the rotation curve in this galaxy.
The emission of the best-fit model is shown overlaid on the p-v diagrams in Fig.\ \ref{fig:j1000}.
In Fig.\ \ref{fig:j1000Channels}, we show the comparison between seven representative channel maps of the $\cii$ emission of J1000+0234 and our best-fit model.
The bottom panels show the residuals of the emission: data$-$model.
It is clear from these comparisons that the $\ciifine$ emission of J1000+0234 is reproduced  very well by an axi-symmetric disc in differential rotation.

\begin{figure}[h]
\centering
  \includegraphics[width=0.5\textwidth]{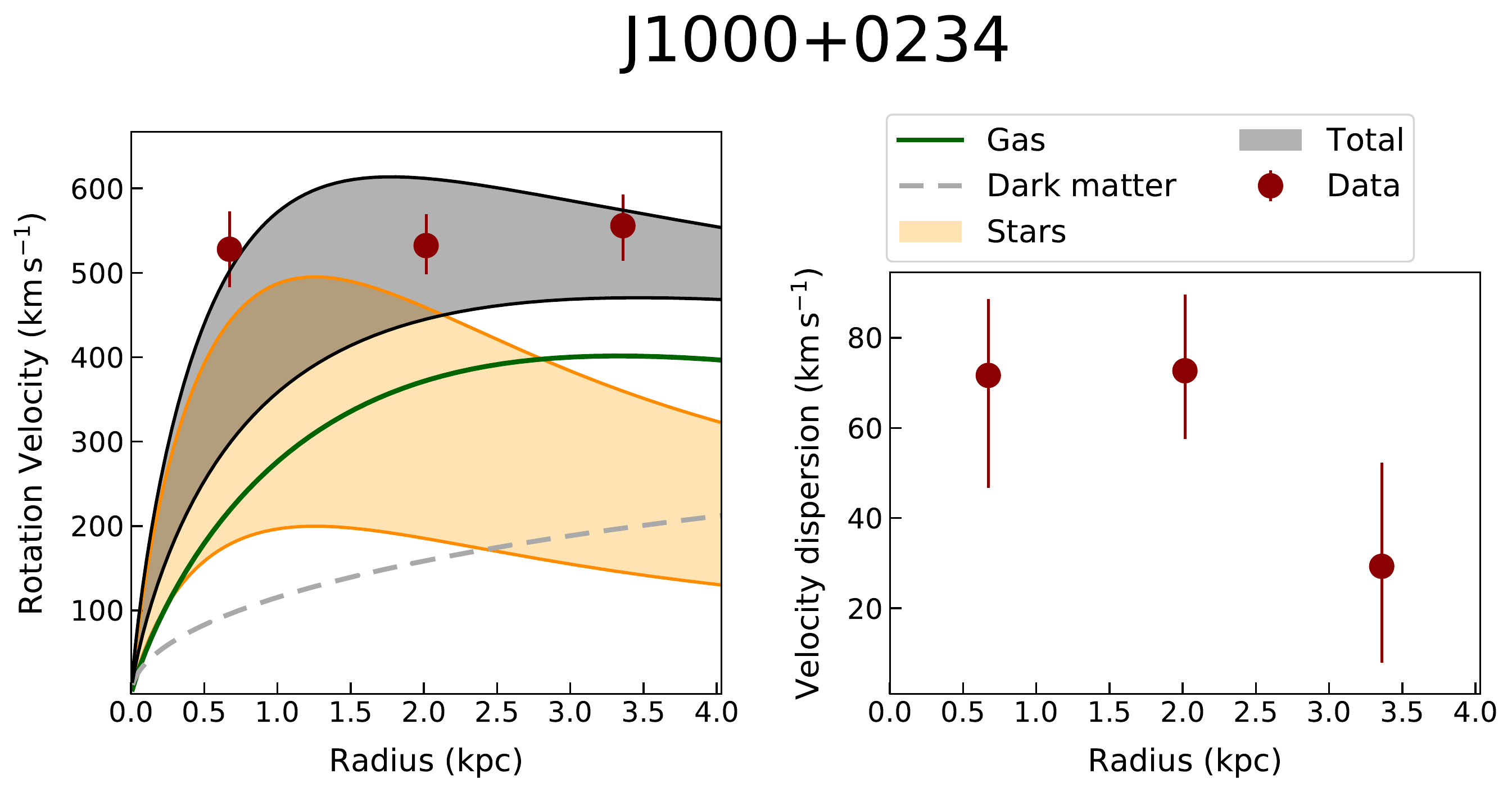}
  \caption{Best-fit rotation velocities and velocity dispersions of J1000+0234. {\em Left panel}: Rotation velocities measured in J1000+0234 (red points) compared to the contributions of different matter components. This comparison is illustrative and not fitted to the data points.
    Orange lines and shades indicate the contribution of the stellar component given the two estimates of stellar mass present in the literature.
    The solid green curve indicates the contribution of the gas disc, which is assumed exponential and is derived from the $\ciifine$ distribution.
    The dashed grey curve indicates the contribution of a dark matter halo given by an NFW profile with a virial mass of $M_{200}=2\times 10^{12}\mo$.
    The black lines and shade indicate the total contribution of all components.
    {\em Right panel}: $\cii$ velocity dispersion measured in J1000+0234.
  }
  \label{fig:j1000VrotVdisp}
\end{figure}

\subsection{Rotation curve analysis of J1000+0234}
\label{sec:rcJ1000}

Given the high inclination of the disc of J1000+0234, the obtained rotation velocities are quite robust and the errors are significantly smaller than for AzTEC/C159.
Figure\ \ref{fig:j1000VrotVdisp} (left panel) shows the measured points and the errors obtained as sums in quadrature of the nominal errors from the fit and the generous variation in inclination of $\pm10\de$.
The rotation curve appears quite flat as the velocities of all the rings are compatible.
We recall that, due to our slight oversampling, these points are not fully independent from one another; however, the fact that the first and the third ring are at the same rotation velocity is a good indication of a flat curve.
The average rotation velocity across the disc is very high, $\vrot=538\pm 40 \kms$, leading to a dynamical mass within the last measured point of $(2.3\pm0.3) \times 10^{11}\mo$.

As for AzTEC/C159, the presence of only three (and not fully independent) points does not allow for a full decomposition of the rotation curve \citep{Cimatti+2019}.
However, calculating the contributions of the various matter components is highly instructive. 
As mentioned, the stellar mass of J1000+0234 is largely unconstrained as there are two very different determinations in the literature.
The stellar size cannot be determined directly since the source is heavily absorbed in the {\em HST} bands and is essentially unresolved by {\em Spitzer} at near-infrared (NIR) wavelengths \citep{Smolcic+2015}.
Therefore, as for AzTEC/C159, we used the FIR effective radius $R_{\rm e}=700 \pc$ \citep{Gomez+2018}, multiplied by 1.4 to convert it to an approximate optical rest-frame effective radius \citep{Fujimoto+2017} and divided by 1.678 to get an exponential scale-length $\Rd$.
This leads, for the two values of stellar masses, to the shaded orange area in Fig.\ \ref{fig:j1000VrotVdisp}.
Thus, the inner point of the rotation curve can be reproduced by a stellar component if the stellar mass is on the high side of the range (Table \ref{tab:j1000}).

For the gaseous disc, we used the $\ciifine$ zeroth moment map to obtain an exponential scale-length $R_{\rm gas}=1.5 \pm0.2 \kpc$ from Galfit \citep{Peng+2010a}.
We calculated the contribution to the rotation curve of this component and found that, in order to reproduce the outer point at $3.3 \kpc$, the gas mass has to be significantly larger than the estimate from \citet{Schinnerer+2008} of $2.8\times 10^{10}\mo$.
This estimate was obtained from $\cofourthree$ using a low $\alpha_{\rm CO}$ factor (0.8 $\mo (\Ktab \kms \pc^2)^{-1}$) that is likely not appropriate for an object such as    J1000+0234 \citep{Papadopoulos+2012, Jimenez+2018}.
The top panel of Fig.\ \ref{fig:j1000VrotVdisp} shows the contribution of a gaseous disc with a mass $\Mg = 1.5\times 10^{11}\mo$.
We note that, as for AzTEC/C159, this is the total mass of the extrapolated exponential profile, which we used to determine the central surface density as $\Mg/(2\pi R_{\rm gas}^2)$.
However, given that the scale-length is quite large ($R_{\rm gas}=1.5 \kpc$), only $\approx 73\%$ of this mass ($\approx 1.1\times 10^{11}\mo$) resides within the extent of the [CII] disc: $R_{\rm ext}\approx 0.6''\approx 4\kpc$.
This mass leads to a ratio between $\cii$ luminosity and gas mass (both in standard units), $\alpha_{\cii}\approx 8$ \citep[using the luminosity $L'$ in ][]{Carilli&Walter2013}, in agreement with values determined for other high-$z$ starbursts \citep{Gullberg+2018}.
We also note that J1000+0234 and AzTEC/C159 having similar gas masses is in line with their having similar dust masses \citep{Smolcic+2015} and $\cii$ fluxes (Tables \ref{tab:aztec} and \ref{tab:j1000}).
All these considerations indicate that the gas mass of J1000+0234 is likely $\approx 1\times 10^{11}\mo$.

The top panel of Fig.\ \ref{fig:j1000VrotVdisp} also shows the contribution from the dark matter halo and the combination of all components (black lines and the grey band).
For the dark matter halo, we have assumed an NFW profile \citep{Navarro+1996} with a concentration $c_{\rm 200}=3.4$ and a mass $M_{200}=2\times 10^{12}\mo$ (see Sect. \ref{sec:rcAztec}).
As for AzTEC/C159, the dark matter is almost negligible in these inner regions of the galaxy for any realistic choice of the virial mass.
We can conclude that the inner kiloparsecs of J1000+0234 probed by our data are fully baryon-dominated, with the gas mass slightly larger than the stellar mass.

The right panel of Fig.\ \ref{fig:j1000VrotVdisp} shows the $\cii$ velocity dispersion in J1000+0234 that we derived with $\bb$.
The values are, as for AzTEC/C159, quite low, leading to a $V/\sigma$ of the order of ten.
However, being the galaxy at high inclination, these velocity dispersions should be considered to be more uncertain than those of AzTEC/C159.
They could in fact be overestimated as some line broadening can occur, along the line of sight, because of projection effects.
This effect is difficult to account for, but it becomes less important in the outer disc; therefore, a realistic value of the velocity dispersion is potentially given by the last ring at $\approx 30 \kms$.
In the following, we attempt to give a better quantification of the velocity dispersion using data at high spectral resolution and focusing on AzTEC/C159.

\subsection{The $\cii$ velocity dispersion in AzTEC/C159}
\label{sec:fitAztecHR}

The $\cii$ disc of AzTEC/C159 is seen at a relatively low inclination with respect to the line of sight.
This makes its rotation velocity quite sensitive to small variations of inclination and therefore relatively uncertain (see error bars in Fig.\ \ref{fig:aztecVrotVdisp}).
However, the gas velocity dispersion is, in these situations, robustly measured.
In order to obtain such a measure, we used $\bb$, which allows us to properly account for beam smearing (see Sect. \ref{sec:fitJ1000}).
We obtained values of about $40\kms$ though with very large uncertainties due to the fact that our data cube (Table \ref{tab:aztecObs}) has a velocity resolution of $27\kms$, which makes it very difficult to reliably measure dispersions around that value. 

\begin{figure*}[ht]
\centering
  \includegraphics[width=0.8\textwidth]{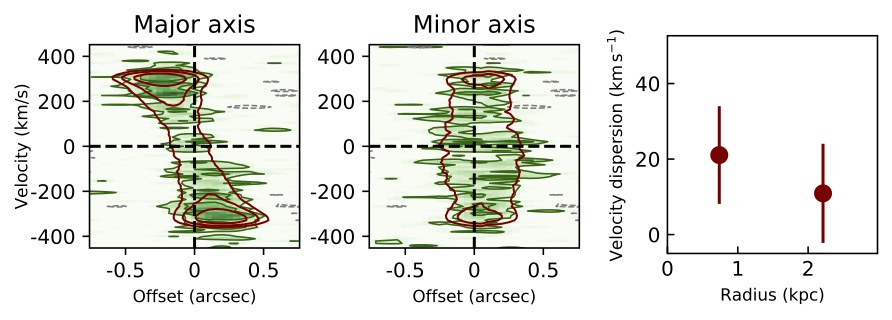}
  \caption{Data and best-fit $\bb$ model of AzTEC/C159 at high spectral resolution.
    {\em Left panel}: Position-velocity diagram along the kinematic major axis of AzTEC/C159 (see Fig.\ \ref{fig:aztec}) obtained using the high spectral resolution data cube (Table \ref{tab:aztecObs}).
    The $\ciifine$ emission in the data is in green shades and contours, while the $\bb$ model is in red contours. Contour levels are at $-2$, 2, 4, and 8 times the r.m.s.\ noise per channel.
    {\em Centre panel}: Position-velocity diagram along the kinematic minor axis displayed analogously to the major-axis plot.
    {\em Right panel}: $\cii$ velocity dispersion obtained from the high spectral resolution data cube.
  }
  \label{fig:aztecHR}
\end{figure*}

We thus produced a high spectral resolution data cube with a channel separation of $10.3 \kms$.
This reduces the S/N per channel but still allows us to obtain a reliable kinematic fit.
Figure\ \ref{fig:aztecHR} shows the emission in this data cube through p-v diagrams along and perpendicular to the major axis, analogous to those shown in Fig.\ \ref{fig:aztec}.
The pattern of rotation is very clear, as in the low spectral resolution data cube, but the emission tends to break up in the middle channels close to systemic velocity due to the low S/N.
The quality of the fit obtained with $\bb$ (shown in red contours) also appears very good at these high spectral resolutions. 
This fit returns an extremely low velocity dispersion, $\sigmag=16\pm 13 \kms$ (averaging the two rings), again of the order of the channel resolution.
These low values of velocity dispersion are robust over any initial guesses for the fit, which we made to vary between $\sigma_{\rm guess}=10 \kms$ and $\sigma_{\rm guess}=300 \kms$.
The values of the rotation velocity obtained with this high spectral resolution data cube are fully compatible with those presented in Sect. \ref{sec:fitAztec}.

\begin{figure}[ht]
\centering
  \includegraphics[width=0.5\textwidth]{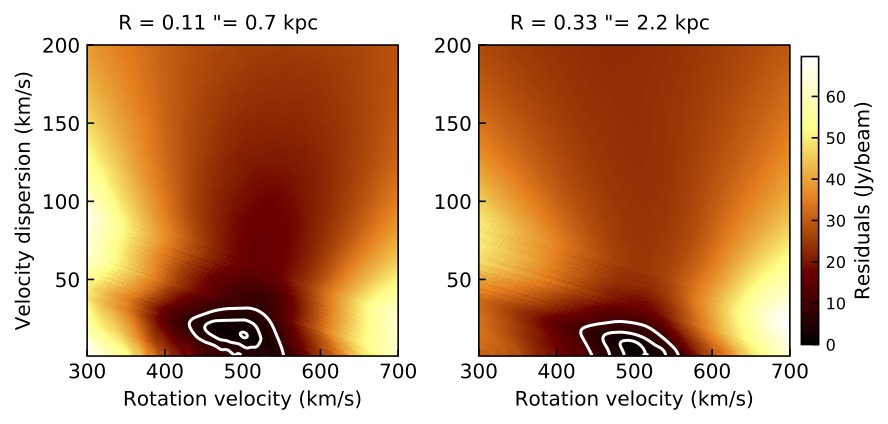}
  \caption{
    Exploration of the large-scale parameter space of the residuals between the data and the models, calculated as the absolute value of model minus data; as data, we used the high spectral resolution data cube of AzTEC/C159, and each model is built with a different pair of $\vrot$ and $\sigmag$, with the other parameters kept fixed.
    {\em Left panel}: Residuals in the inner ring; contours show levels where the residuals exceed 2, 5, and 10 times the minimum value, respectively.
    {\em Right panel}: Same, but for the outer ring.
  }
  \label{fig:spaceparHR}
\end{figure}

To further test the reliability of the above fit, we ran $\bb$ with the {\sf spacepar} keyword turned on.
This keyword allows the user to explore the full parameter space of two parameters by calculating the residual (|model$-$data|) in a grid with arbitrary spacing.
 We used as free parameters the rotation velocity ($\vrot$) and the velocity dispersion ($\sigmag$) and explored a large portion of the space around the best-fit values determined above.
This exploration was done for the two rings separately, and the final result is shown in Fig.\ \ref{fig:spaceparHR}.
There we see the values of the residual for any pair of $\vrot$ and $\sigmag$ in a range of about $200 \kms$ in each direction.
It is clear that the exploration of both rings shows a well-defined absolute minimum, and there is no sign of a secondary minimum. 
The contours show regions where the residuals are at two, five, and ten times the absolute minimum.
The favoured models are clearly at $\vrot\approx 500 \kms$ and $\sigmag \lsim 20 \kms$, in agreement with the best-fit values determined above.

We also ran kinematic fits of our two galaxies using Galpak$^{\rm 3D}$ \citep{Bouche+2015}.\ We found values for the rotation velocities, the velocity dispersions, and the inclinations that are fully consistent, within the errors, with the $\bb$ runs.

\section{Discussion}
\label{sec:discussion}

In the previous sections, we have seen the results of 3D kinematic fits of the ALMA $\ciifine$ data cubes of two starburst galaxies at $z\approx 4.5$ obtained with the software $\bb$.
These two galaxies, which show a number of similarities with each other based on their photometric properties (see Tables \ref{tab:aztec} and \ref{tab:j1000}), also appear very similar in terms of dynamical properties as estimated from the $\cii$ gas kinematics.
The general picture that we obtain is of very fast rotating gas discs with relatively low velocity dispersion as well as $V/\sigma$ values comparable to $z=0$ spiral galaxies.
In the following sections, we discuss the reliability of our results as well as possible evolution scenarios for these galaxies that emerge from our dynamical analysis.

\subsection{Determining if velocity gradients can indicate anything other than rotation}

In Sects. \ref{sec:fitAztec} and \ref{sec:fitJ1000}, we have seen how well a model of a rotating disc can reproduce the emission in the data cubes of both AzTEC/C159 and J1000+0234.
We wanted to determine whether this is a unique possibility or if there are other configurations that can give a similar pattern.
In general, a velocity gradient seen at low angular resolution should be taken with caution.
At high angular resolution, a rotating disc gives an extremely distinctive pattern that leaves no room for alternative interpretations \citep{Swaters+2002, Fraternali+2002, Walter+2008}.
However, velocity gradients can also be evidence of outflows \citep{Martini+2018}, mergers \citep{Sweet+2019}, or, potentially, gas inflow.
Thus, in particular at high redshift, one should question whether or not these phenomena can play a role.

However, a careful inspection of our data, looking both at the channel maps and the p-v diagrams, reveals the distinctive aspect of our velocity gradients.
The channel maps of both galaxies show a remarkably clear pattern of a rotating disc with a flat rotation curve as the emission accumulates at the extreme (approaching and receding) channels (see panels 2 and 6, from left to right, in Figs. \ref{fig:aztecChannels} and \ref{fig:j1000Channels}), becoming dimmer instead close to systemic velocity.
This is also plainly visible in the p-v diagrams in Figs.\ \ref{fig:aztec} and \ref{fig:j1000} and especially in Fig.\ \ref{fig:aztecHR}, where we also see, by looking at the model contours, that a rotating disc predicts precisely this pattern.
Moreover, the relatively low inclination of AzTEC/C159 allows us to admire another distinctive feature of rotating discs: the `butterfly' opening of the channel maps close to the systemic velocity (Fig.\ \ref{fig:aztecChannels}).
Finally, both galaxies show a velocity gradient nicely aligned with the elongation of the $\cii$ emission (i.e.\ the morphological and kinematic axes coincide), a further feature of rotating discs.
We conclude that there is no reason to doubt that the structures that we are observing in these two galaxies are rotating discs at $z\approx4.5$.
This is in line with recent results obtained for similar galaxies at slightly higher angular resolution \citep{Neeleman+2020, Lelli+2021} with the help of gravitational lensing \citep{Rizzo+2020}.

\subsection{Understanding the low gas turbulence in AzTEC/C159}

The gas velocity dispersion of AzTEC/C159 (Sect. \ref{sec:fitAztecHR}) is significantly lower relative to values determined by other authors for similar systems \citep[e.g.][]{Hodge+2012, Carniani+2013} as well as expectations from numerical simulations \citep{Zolotov+2015,Pillepich+2019}.
From the observational side, our result is a combination of two improvements with respect to other works as 1) we used $\bb$ to fully account for the beam smearing and 2) we used a data cube at very high spectral resolution ($\approx 10\kms$).
From the theoretical side, it is useful to estimate whether such a low dispersion can be expected from simple arguments.

We assumed that the velocity dispersion that we measure for the $\cii$ is representative of the cold molecular ISM of AzTEC/C159, which has a total mass $\Mg\simeq 1.5\times 10^{11} \mo$ (Table \ref{tab:aztec}).
Given the SFR of this galaxy of $740 \moyr$, one would expect a supernova rate (SNR) of ${\rm SNR} \approx 7\yrinv$, assuming a standard initial mass function.
The energy input from supernovae that can feed turbulence is then
\begin{equation}
  \dot E_{\rm SN} \simeq 2.2 \times 10^{43} \left(\frac{\eta}{0.1}\right)
  \left(\frac{{\rm SNR}}{7\yrinv}\right) \ergs,
  \label{eq:energy}
\end{equation}
where we have taken the standard total energy per supernova of $10^{51}\erg$ and an efficiency of transferring kinetic energy to the ISM of 10\% \citep{Chevalier1974,Ohlin+2019}.
The total turbulent (kinetic) energy of the gas disc of AzTEC/C159 can simply be estimated as
\begin{equation}
  E_{\rm turb}=\frac{3}{2}\Mg\sigmag^2, 
\end{equation}
where $\sigmag$ is the gas velocity dispersion that we can expect to measure and the factor 3 comes from the fact that this dispersion is measured along the line of sight.
Turbulence is dissipated on timescales of $\tau \approx h/\sigmag$ \citep{MacLow1999}, where $h$ is the scale-height of the disc that unfortunately is unknown and we assume to be in the range $h=100$-$500\pc$.
To maintain turbulence at the observed value, supernovae must provide energy at the rate $\dot E_{\rm turb}=E_{\rm turb}/\tau$.
By equating this energy input \citep[see also][]{Rizzo+2020} with the one available in Eq.\ (\ref{eq:energy}), we can estimate a gas velocity dispersion of
\begin{equation}
\sigmag\approx 31 \left(\frac{\eta}{0.1}\frac{\rm SNR}{7\yrinv}\frac{h}{200\pc}\right)^{1/3}
\left(\frac{\Mhtwo}{1.5\times 10^{11} \mo}\right)^{-1/3}\kms,
\label{eq:sigmaTheo}
\end{equation}
and, considering the range in $h$ and the errors in the SNR (i.e.\ the SFR) and $\Mhtwo$, we obtain that $\sigmag$ is in the range 22-$50\kms$ for an efficiency $\eta=0.1$.
In conclusion, the values of gas velocity dispersion that we have measured are on the low side of this estimate.
They could point to an overestimate of the SFR or a relatively low efficiency of energy transfer to the ISM (of a few percent).
Recent similar estimates, carried out in local galaxies, indeed show that only a few percent of the supernova energy goes to the feeding of the ISM turbulence \citep{Bacchini+2020}.

In this context, it is important to point out that a velocity dispersion of $\approx 20-30 \kms$, although lower than other literature determinations, is still quite large considering that the $\ciifine$ line is tracing a rather cold component of the ISM.
In local galaxies, molecular gas has typical velocity dispersions of $\approx 5 \kms$ \citep{Marasco+2017, Koch+2019, Bacchini+2020}, while atomic neutral gas has $\approx 10 \kms$ \cite[e.g.][]{Tamburro+2009, Bacchini+2019}.
Thus the gas velocity dispersions that we are measuring in our high-$z$ starburst galaxies are higher than those of $z=0$ discs, but not exceedingly so.

In the above analysis, we have focused on AzTEC/C159, which gives the most reliable measure of the velocity dispersion.
The high inclination of J1000+0234 is, instead, likely broadening the line profiles, an effect that $\bb$ can only partially compensate for \citep{DiTeodoro&Fraternali2015}.
Moreover, the large uncertainties in SFR and gas mass (Table \ref{tab:j1000}) would make the above calculation not very predictive since Eq.\ (\ref{eq:sigmaTheo}) returns values in the range $14-100 \kms$ for $\eta=0.1$.

\subsection{Evolution of high-$z$ starbursts}

Massive starburst galaxies at $z\approx 4$ are believed to be the progenitors of massive ETGs \citep[e.g.][]{Toft+2014}.
This expectation is mostly based on the extremely high SFRs combined with high stellar and gas masses.
The depletion time of these galaxies is typically significantly less than 1 Gyr, pointing to the possibility that the whole stellar population can be formed before, say, $z\approx 2-3$ and that the galaxy will eventually evolve passively and through dry mergers.
Massive ETGs at $z\approx 2$ are indeed observed, with stellar masses comparable to local ETGs and rather compact sizes \citep[e.g.][]{Cimatti+2008, Cappellari2016}.
Our results allow us to test this evolutionary hypothesis dynamically.

\begin{figure}[ht]
\centering
  \includegraphics[width=0.5\textwidth]{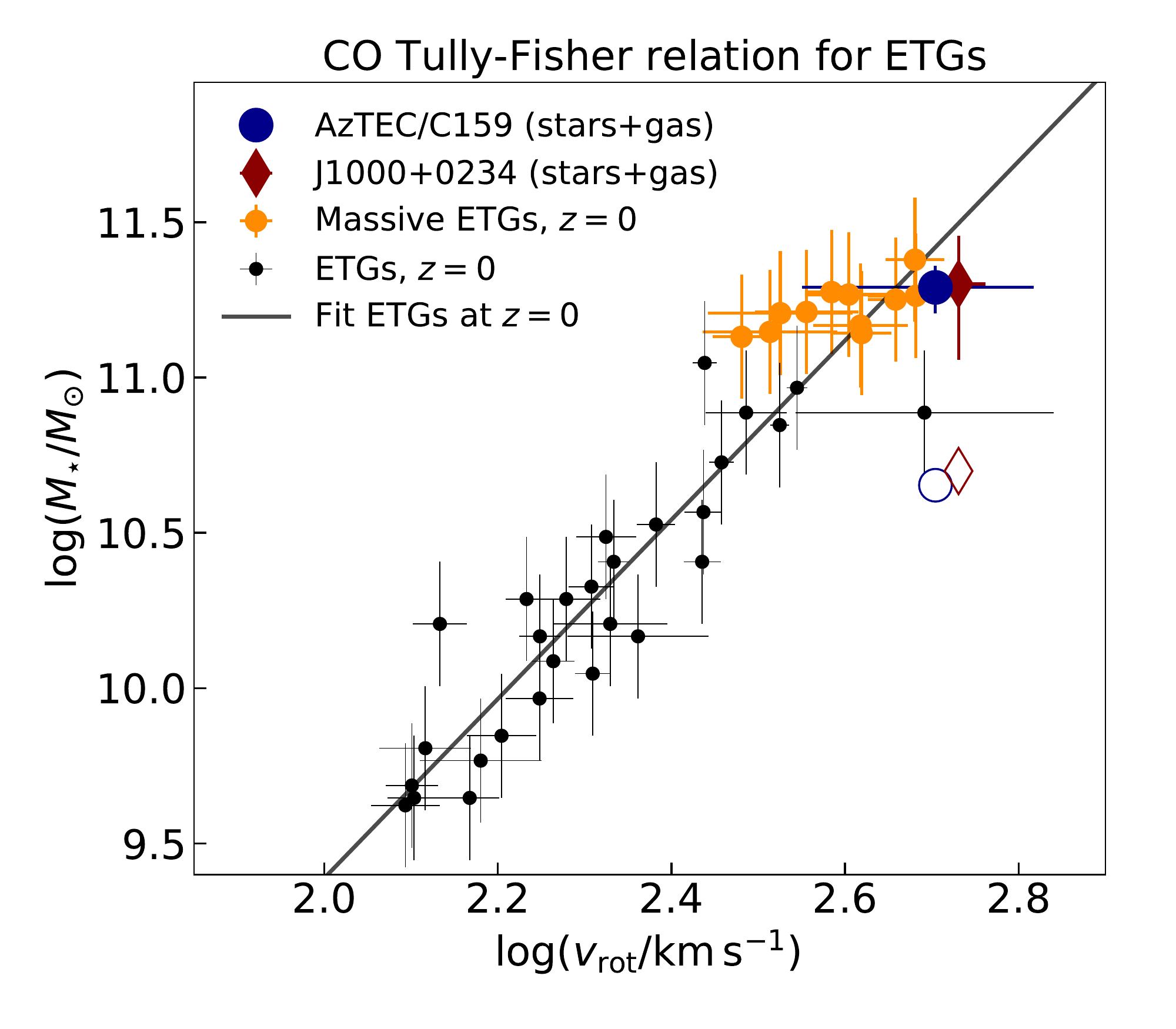}
  \caption{
    Stellar-mass Tully-Fisher (analogue) relation for ETGs at $z=0$ from \citet{Davis+2016} compared with the positions of our starburst galaxies at $z\approx4.5$.
    In the local ETGs, the rotational speed is derived from the inner CO discs, while in our galaxies it is obtained from the $\cii$ line.
    The empty symbols show the positions of our galaxies based on the observed stellar mass, while the filled symbols show their positions assuming that all the observed gas mass is converted into stars.
  }
  \label{fig:beastsTFR}
\end{figure}

The kinematics and dynamics of local ETGs are typically studied using stellar absorption lines \citep{Thomas+2007, Cappellari+2013, Corsini+2017, Zhu+2018}.
However, some ETGs have cold gas that allows a variety of kinematic studies \citep{Davis+2011a,Serra+2014, Shelest&Lelli2020}.
In particular, it is relatively common for ETGs (roughly 20\% of cases) to have central compact discs of molecular gas with typical sizes of $\sim 1\kpc$ that allow us to probe the gravitational potential in the inner regions \citep{Davis+2013}.
These discs often show regular rotation patterns displayed by the global profiles \citep{Davis+2011b} and by resolved observations \citep{Young+2005, North+2019}.
It has been shown \citep{Davis+2016} that these CO discs fulfil the ETG analogue of a stellar-mass Tully-Fisher relation \citep{Tully&Fisher1977}.
We show this relation in Fig.\ \ref{fig:beastsTFR} for the sample of galaxies in \citet{Davis+2016}.
We note that the velocities were slightly revised due to better determinations of the inclinations of the discs (T.\ Davis, private comm.) and that the masses are determined from the $K$-band luminosity assuming a mass-to-light ratio M/L$=0.5{M}_{\odot}/{\rm L}_{\odot,K}$.
In the same figure, we also include our two high-$z$ starbursts.
For their rotation velocities, we used the values that we obtained from their $\cii$ discs, which are similar in size to the CO discs in local ETGs.
For the stellar masses, we used the estimated stellar mass (empty symbols), and we estimated their `final' mass by assuming that they will convert all their gas into stars (filled symbols).
For J1000+0234, we used an average stellar mass between the two estimates (Table \ref{tab:j1000}) and the gas mass that we determined through the rotation curve decomposition (Sect. \ref{sec:rcJ1000}).
After the conversion of gas into stars, our two high-$z$ starbursts appear to fully overlap with the local ETGs.

A number of considerations can be drawn from the comparison presented in Fig.\ \ref{fig:beastsTFR}.
First, it appears that the inner potential well of our galaxies at $z\approx 4.5$, quantified by their circular speed, is indeed comparable with that of the most massive ETGs that we observe today.
Second, the conversion of the gas mass into stellar mass, which is expected to occur in a few hundred megayears (depletion time), can make these galaxies lie in the same scaling relation as massive local ETGs.
We note that although we are unable to give a precise quantification of the stellar effective radius of our galaxies, it is likely that they will be smaller than those of local ETGs that have the same stellar mass \citep{vdWel+2014, Cappellari2016}, as was shown by \citet{Gomez+2018} using the FIR sizes.
Thus, the effective radius of these galaxies will likely grow with time, which will also have potential consequences on the precise value of the inner circular speed.
However, in general terms, we can envisage a scenario in which our galaxies convert all their gas into stars, go through a quenching phase \citep[e.g.][]{Peng+2010b}, and then evolve passively and through mergers to be observed today at the massive end of the ETGs.

The most puzzling aspect of the above scenario is the extremely fast rotation velocity that we measured.
On the one hand, there is no surprise that the circular speed in these high-$z$ starbursts is $500 \kms$ because it is also the case for the most massive local ETGs.
On the other hand, however, we observe a very large mass ($\sim 10^{11}\mo$) of gas rotating at that speed; when this gas is turned into stars, it is conceivable to imagine that these stars will rotate very fast.
Thus, we should end up with an extremely massive and compact rotation-supported stellar disc.
Stellar discs with similar properties have been found in a some gravitationally lensed galaxies at $z\approx 2-2.5$ \citep{Newman+2015, Toft+2017}, and it is possible that the two starbursts studied in this paper are progenitors of this kind of object.
However, it is important to note that the $V/\sigma$ values of the $z\approx 2$ stellar discs are much smaller than what we measured in our $z\approx 4.5$ starbursts.

In the local Universe, stellar discs rotating at $v>300 \kms$ are extraordinarily rare  \citep{Posti+2018, Ogle+2019}, and it is unlikely that they are the end point of the evolution of high-$z$ starburst galaxies, given the reasons mentioned above.
Therefore, it appears necessary that the stars that form in these high-$z$ fast rotating and cold discs redistribute into a pressure-supported system analogous to a slow rotator ETG \citep{Cappellari2016}.
A phenomenon that can help in this direction is a major merger or a large number of minor mergers \citep[see discussion in][]{Rizzo+2020}.
However, it is tempting to also consider the possibility of global disc instabilities \citep[e.g.][]{Ostriker&Peebles1973,Sellwood2013}.
We leave the exploration of these possibilities to future works.

\section{Conclusions}
\label{sec:conclusions}

We have studied the kinematics of two starburst galaxies at $z\approx 4.5$ (AzTEC/C159 and J1000+0234) using the $\ciifine$ emission line observed with ALMA at an angular resolution of $\approx 0.3''$, which corresponds to $\approx 2 \kpc$.
We have modelled the $\cii$ emission using the code $\bb$, which fits a tilted-ring galaxy model in 3D space by producing artificial data cubes that include all observational biases.
In particular, $\bb$ is capable of accounting for the effect of beam smearing, which is particularly severe with data at these angular resolutions.
The main results of this work are as follows.

\begin{enumerate}

\item
  The $\ciifine$ emission of our two galaxies is perfectly reproduced by models of gas discs in regular (differential) rotation.
\item
  The best-fit rotation velocities are $506^{+151}_{-76} \kms$ and $538\pm40 \kms$ for AzTEC/C159 and J1000+0234, respectively.
  The errors in the former galaxy are larger due to the relatively low inclination of the disc.
\item
  We show that the masses of the stellar and gaseous disc components needed to reproduce these rotation velocities are compatible with the literature values for AzTEC/C159, while J1000+0234 requires higher values of $\approx 10^{11}\mo$.
  Both galaxies are strongly baryon-dominated in the probed regions (the inner few kiloparsecs) with a negligible dark matter contribution.
\item
  The best-fit gas (intrinsic) velocity dispersions are at least a factor of ten lower than the rotation velocities.
  As these values are of the order of the velocity resolution of our initial data cube, we repeated the kinematic fit for AzTEC/C159 in a data cube with a spectral resolution of $10\kms$, finding a best-fit velocity dispersion $\sigmag = 16\pm 13 \kms$, which leads to $V/\sigma \gsim 20$.
\item
  We showed that the surprisingly low velocity dispersion of AzTEC/C159 can be compatible with turbulence fed by supernovae with an efficiency $\eta\lsim 10\%$.
\item
  Our galaxies appear to have potential wells (demonstrated by the measured circular speeds) that are very similar to those of local massive ETGs.
  A comparison of their positions in the ETG analogue of the stellar-mass Tully-Fisher relation, obtained from the rotation of CO discs in ETGs, reveals an almost perfect overlap provided that all the observed gas in our high-$z$ galaxies is converted into stars.
\end{enumerate}  

In conclusion, our results provide dynamical support to the idea that high-$z$ starbursts are progenitors of local ETGs.
If the galaxies already have extremely deep potential wells at $z\approx 4.5$, the conversion of gas into stars, combined with the subsequent quenching of star formation, should produce objects that can then both structurally and dynamically evolve passively into ETGs.
The puzzle that remains unsolved concerns the transformation of a highly rotation-supported system into a likely slow rotator ETG.

\begin{acknowledgements}
  We thank Timothy Davis for providing his data on local ETGs, Eric Emsellem for stimulating discussions and Ian Smail for helpful comments.
  We also thank an anonymous referee for an useful and constructive report.
  FF acknowledges support from the Friedrich Wilhelm Bessel Research Award Programme of the Alexander von Humboldt Foundation.
  AK and BM acknowledge support by the Collaborative Research Council 956, sub-project A1, funded by the Deutsche Forschungsgemeinschaft (DFG).
  C.G.G.\ acknowledges support from the European Research Council (ERC) Consolidator Grant funding scheme (project ConTExt, grant number: 648179).
  This work makes use of the following ALMA data: 2012.1.00978.S and 2015.1.01564.S.
  ALMA is a partnership of ESO (representing its member states), NSF (USA), and NINS (Japan), together with NRC (Canada), NSC and ASIAA (Taiwan), and KASI (Republic of Korea), in cooperation with the Republic of Chile.
  The Joint ALMA Observatory is operated by ESO, AUI/NRAO, and NAOJ.
\end{acknowledgements}

\bibliographystyle{aa} 
\bibliography{fraternali} 

\end{document}